\documentclass{aa}

\usepackage[varg]{txfonts}

\defcitealias{dae13}{D13}

\begin{document}

\title{The frequency of accretion disks around single stars: Chamaeleon~I}

\titlerunning{Disks Around Single Stars}
\authorrunning{Daemgen et al.}
\author{Sebastian Daemgen\inst{\ref{inst1}}
\and R. Elliot Meyer\inst{\ref{inst1}}
\and Ray Jayawardhana\inst{\ref{inst2}}
\and Monika G. Petr-Gotzens\inst{\ref{inst3}}}

\institute{Department of Astronomy \& Astrophysics, University of Toronto, 50 St. George Street, Toronto, ON, Canada M5H 3H4; \email{daemgen@astro.utoronto.ca}\label{inst1}
\and York University, Faculty of Science, 4700 Keele Street, Toronto, ON M3J 1P3, Canada\label{inst2}
\and European Southern Observatory, Karl-Schwarzschildstr.\ 2, 85748, Garching, Germany\label{inst3}}

\abstract
{It is well known that stellar companions can influence the evolution of a protoplanetary disk. Nevertheless, previous disk surveys did not -- and could not -- consistently exclude binaries from their samples.}
{We present a study dedicated to investigating the frequency of ongoing disk accretion around \emph{single} stars in a star-forming region.}
{We obtained near-infrared spectroscopy of 54 low-mass stars selected from a high-angular resolution survey in the 2--3\,Myr-old Chamaeleon\,I region to determine the presence of Brackett-$\gamma$ emission, taking the residual chance of undetected multiplicity into account, which we estimate to be on the order of 30\%. The result is compared with previous surveys of the same feature in binary stars of the same region to provide a robust estimate of the difference between the accretor fractions of single stars and individual components of binary systems.}
{We find Br$\gamma$ emission among $39.5^{+14.0}_{-9.9}$\% of single stars, which is a significantly higher fraction than for binary stars in Chamaeleon\,I. In particular, close binary systems with separations $<$100\,AU show emission in only $6.5^{+16.5}_{-3.0}$\% of the cases according to the same analysis. {The emitter frequency of wider binaries appears consistent with the single star value. Interpreting Br$\gamma$ emission as a sign of ongoing accretion and correcting for sensitivity bias, we infer an accretor fraction of single stars of $F_\mathrm{acc}=47.8^{+14.0}_{-9.9}$\%. This is slightly higher but consistent} with previous estimates that do not clearly exclude binaries from their samples.}
{Through our robust and consistent analysis, we confirm that the fraction of young single stars harboring accretion disks is much larger than that of close binaries at the same age.  Our findings have important implications for the timescales of disk evolution and planet formation.}


\maketitle

\section{Introduction}
The observed high abundance of stars with protoplanetary disks in the youngest star-forming regions suggests that most, if not all, low-mass stars undergo the T Tauri phase during the first few Myrs of their evolution. The fraction of actively accreting stars subsequently evolves to essentially 0\% within $\sim$10\,Myr \citep[e.g.,][]{jay06,fed10}, setting an upper limit to the time available for the formation of gas giant planets. As the decline of accretion disk frequency appears to be gradual, i.e., not all stars lose their gas disk at the same age, this implies that gaseous planets may form on even shorter timescales of $\lesssim$1\,Myr, in agreement with disk fragmentation theories \citep[e.g.,][]{bos97}.

A qualitatively similar disk frequency evolution has been observed for stars bound in binary and higher-order multiple systems, however, with the notion that the frequency of disks is on average lower than in single stars \citep{dae13,kra12b}. This may be explained by the smaller dynamically allowed size of the disks \citep[truncation of the disk to $\sim$1/3 of binary separations,][]{art94}, and the accordingly smaller mass reservoir \citep{har12} while mass accretion rates are indistinguishable from those of single stars \citep[][Daemgen et al.\ 2015, \emph{in prep.}]{whi01}. 

A quantitative analysis of the magnitude of the effect of binarity on disk dispersal timescales has been hindered by two factors. On one hand, different disk tracers (e.g., IR colors, optical/IR line emission) are known to return different disk frequencies as they probe different temperatures, radii, and processes that are not necessarily correlated. Thus, a useful comparison between the disk frequency of binaries and single stars can only be made when using the same disk tracer. The facts that only spatially resolved binary stars can eliminate dependencies on, e.g., stellar mass, and that the most significant impact of binarity on disk evolution has been observed for close binary systems with separations $\lesssim$100\,AU \citep{kra12b,dae13} restrict studies to high-angular resolution observations, mostly with adaptive optics in the near-infrared. Such data are available for a limited number of binary stars \citep{pra03,mon07,kra12b,dae12a,dae13}. Suitable observations of single stars with similar techniques, which would allow unbiased comparison, are rare.

On the other hand, existing ``single star'' disk frequency estimates do not account for the high chance of undetected binary stars contaminating their samples. As the disk frequency of binaries is lower than that of singles, the typically quoted frequencies will be biased toward lower fractions and the actual disk lifetime is veiled. In an attempt to account for this effect statistically, \citet{kra12b} corrected the observed disk frequencies for the frequency of binary stars for a dozen regions: They found a consistently large disk fraction until an age of $\sim$3\,Myr and a subsequent sudden drop consistent with some disk evolution models \citep{ale09}. While providing possible evidence for a qualitatively different disk frequency evolution than previously assumed, the fact that \citet{kra12b} combine a variety of dust and accretion tracers and a statistical correction instead of a direct frequency measurement limits the applicability of the results.

The present study investigates the presence of accretion in single stars in the Chamaeleon\,I (Cha\,I) star-forming region. 
By using the same accretion-sensitive spectroscopic feature as a previous study of binary stars in the same region \citep[Brackett-$\gamma$;][subsequently \citetalias{dae13}]{dae13}, this will be the first study to unambiguously compare the single and binary star accretion fractions. 
Sect.~\ref{sec:data} describes the sample selection, observations and data reduction, Sect.~\ref{sec:results} presents the results. We discuss the implications in Sect.~\ref{sec:discussion} and conclude in Sect.~\ref{sec:conclusion}.

\section{Observations \& data reduction}\label{sec:data}
\subsection{Sample selection}\label{sec:biases}
All targets are confirmed members of the 2--3\,Myr-old Cha\,I star-forming region at a distance of $\sim$160\,pc \citep{luh07,whi97}, {selected from the high-angular resolution survey by \citet{laf08} to exclude binaries in a separation range between 0\farcs05 and 5\arcsec\ with a contrast of $\Delta K_\mathrm{s}\approx3$\,mag or brighter. We further checked the literature {for recent companion discoveries and} spectroscopic binary stars in the sample and deselected these \citep{ngu12,joe08,sch13}.} In order to serve as a comparison sample for the binary star observations by \citetalias{dae13}, target stars were selected to have a similar sample structure. This includes magnitude cuts at R$<$17\,mag and 6$<$$K$$<$13.0\,mag, and a similar spatial distribution in the Cha\,I region. {To ensure a homogeneous analysis, we limit the spectral type range to lie between K4 and M7 where coverage in the binary sample (\citetalias{dae13}) is best.} 
{From a total of 68 target candidates, 54 have been observed}. 

\begin{figure}
\centering
\includegraphics[width=\columnwidth]{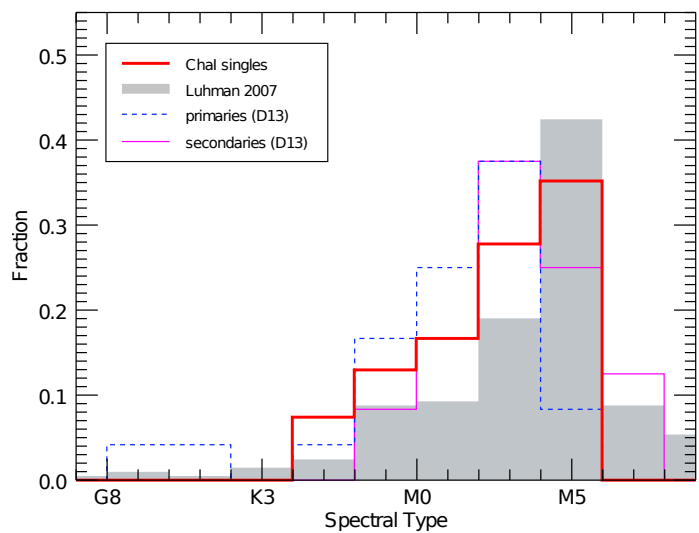}
\caption{\label{fig1}Spectral type distribution of the Cha\,I single star sample. For comparison, we also show the overall distribution of Cha\,I members \citet{luh07} as well as components of binary stars in \citetalias{dae13}.}
\end{figure}

As the source catalog \citep{laf08} surveyed essentially all members of Cha\,I brighter than K=13\,mag known at the time, and since our sample selection is random within our selection criteria, we expect our target sample to represent all cluster members in the same magnitude and spectral type range. The observed targets are distributed within $\sim$1$^\circ$ from the cluster center, an area which is host to $\sim$92\% of all members identified by the large census of \citet{luh07}. Of the 40 targets that are part of the \emph{Spitzer Space Telescope} census by \citet{luh08b}, 24 were classified as infrared excess sources (class II) and 17 were found without excess (class III). The class II fraction of our subsample is 58$^{+7}_{-8}$\%, consistent with the overall class II fraction in Cha\,I of $\sim$50\% \citep{luh08b}. 
Fig.~\ref{fig1} shows the spectral type distribution of the sample. K-S tests are used to compare our target sample with components of binary stars \citepalias{dae13} and the \citeauthor{luh07} census in the same spectral type range. While our target sample appears indistinguishable ($P=0.6$) from the less massive components of binaries -- by definition primary stars of binaries are of earlier type than average -- we find marginal evidence that our sample does not share the same underlying spectral type distribution with stars in \citet[$P=0.05$]{luh07}.
The median spectral types of our sample and the \citeauthor{luh07} sample are, however, different by only one subclass (M2.8 and M3.8, respectively) and the shape of the distribution overall similar.

In addition, most source parameters (spatial distribution, luminosities, ages) are also consistent with the set of binary stars presented by \citetalias{dae13}. By definition, primary stars have on average higher masses than secondaries. The spectral type and effective temperature distributions of our sample are more closely followed by those of the secondaries, while primaries have on average $\sim$2 subclasses earlier spectral types. Given the overall limited spectral type range of this survey and since most relevant parameters (such as accretion) are not a steep function of spectral type, we treat all binary components, irrespective of their primary or secondary nature, as one sample to ensure reliable statistics.

Table~\ref{tab1} lists our target stars together with spectral types, 2MASS $K_\mathrm{s}$-band magnitudes, and extinction measurements from the literature \citep{cut03,luh04}. 

\begin{table*}
\caption{Target star data\label{tab1}}
\centering
\tiny
\begin{tabular}{lcccr@{$\pm$}lccc}
\hline\hline
  &
 {Spectral} &
 {$K$\tablefootmark{c}} &
 {Av\tablefootmark{b}} &
 \multicolumn{2}{c}{EW(Br$\gamma$)\tablefootmark{d}} &
 K-type& 
 &
 \\
 {Identifier\tablefootmark{a}}  &
 {Type\tablefootmark{b}} &
 {(mag)} &
 {(mag)} &
 \multicolumn{2}{c}{(\AA)} &
 STD\tablefootmark{e} &
 P$_\mathrm{em}$\tablefootmark{f} &
 Survey\tablefootmark{g}\\
 \hline
\object{CHSM 1715} 		& M4.25	&  10.9 & 2.6 	&     0.19&  0.35 &  *	 &  0.00 &  IM   	\\ 
\object{CHX 18N}		& K6	&  7.77 & 0.12 	&  $-$0.36&  0.18 &  *	 &  0.06 &  IM,RV	\\ 
\object{CHXR 14N}		& K8	&  9.60 & 0.78 	&     0.29&  0.20 &  *	 &  0.00 &  IM,RV	\\ 
\object{CHXR 20}		& K6	&  8.88 & 3.9 	&  $-$0.31&  0.20 &  *	 &  0.02 &  IM,RV	\\ 
\object{CHXR 21}		& M3	&  9.66 & 3.2 	&  $-$0.14&  0.29 &  	 &  0.00 &  IM,RV	\\ 
\object{CHXR 22E}		& M3.5	&  10.0 & 6.5 	&  $-$0.25&  0.24 &  *	 &  0.00 &  IM   	\\ 
\object{CHXR 33}		& M2.5	& \ldots& 2.2 	&  $-$0.03&  0.15 &  *	 &  0.00 &  IM   	\\ 
\object{CHXR 35}		& M4.75	&  10.9 & 1.1	&  $-$0.48&  0.31 &  	 &  0.01 &  IM   	\\ 
\object{CHXR 48}		& M2.5	&  9.80 & 1.4 	&     0.09&  0.27 &  	 &  0.00 &  IM,RV	\\ 
\object{CHXR 54}		& M1	&  9.50 & 0.59 	&     0.55&  0.15 &  	 &  0.00 &  IM,RV	\\ 
\object{CHXR 55}		& K4.5	&  9.29 & 1.1 	&  $-$0.38&  0.23 &  	 &  0.02 &  IM,RV	\\ 
\object{CHXR 57}		& M2.75	&  10.0 & 0.67	&     0.37&  0.21 &  *	 &  0.00 &  IM,RV	\\ 
\object{CHXR 84}		& M5.5	&  10.8 & 0.00	&  $-$0.69&  0.28 &  	 &  0.11 &  IM   	\\ 
\object{ESO-HA 553}		& M5.6	&  11.5 & 0.00	&  $-$0.73&  0.28 &  	 &  0.13 &  IM   	\\ 
\object{Hn 2}			& M5	&  9.99 & 1.9 	&  $-$0.16&  0.41 &  	 &  0.00 &  IM   	\\ 
\object{Hn 5}			& M4.5	&  10.1 & 1.9	&  $-$1.73&  0.21 &  	 &  1.0  &  IM,RV	\\ 
\object{Hn 7}			& M4.75	&  11.0 & 1.1	&  $-$0.31&  0.32 &  	 &  0.00 &  IM   	\\ 
\object{Hn 11}			& K8	&  9.44 & 7.1 	&  $-$1.37&  0.21 &  *	 &  1.0  &  IM   	\\ 
\object{Hn 17}			& M4	&  11.2 & 0.48 	&     0.10&  0.30 &  	 &  0.00 &  IM,RV	\\ 
\object{ISO-ChaI 52}		& M4	&  10.6 & 1.1 	&     0.10&  0.51 &  	 &  0.00 &  IM,RV	\\ 
\object{ISO-ChaI 76}		& M2.75	&  10.3 & 17	&  $-$0.21&  0.20 &  	 &  0.01 &  IM   	\\ 
\object{ISO-ChaI 80}		& M3	&  9.39 & 6.0	&     0.30&  0.25 &  *	 &  0.00 &  IM,RV	\\ 
\object{ISO-ChaI 99}		& M4.5	&  10.3 & 0.00	&     0.16&  0.40 &  *	 &  0.00 &  IM   	\\ 
\object{ISO-ChaI 189}		& M1.25	&  8.67 & 20	&  $-$2.50&  0.23 &  *	 &  1.0  &  IM   	\\ 
\object{ISO-ChaI 196}		& K7  	&  9.16 & 19   	&     0.07&  0.15 &  	 &  0.00 &  IM   	\\ 
\object{ISO-ChaI 237}		& K5.5	&  8.62 & 6.8 	&  $-$0.48&  0.21 &  	 &  0.08 &  IM,RV	\\ 
\object{ISO-ChaI 250}		& M4.75	&  10.7 & 6.0  	&     0.01&  0.58 &  *	 &  0.00 &  IM   	\\ 
\object{ISO-ChaI 274}		& M4.5	&  10.7 & 1.0 	&  $-$0.13&  0.34 &  *	 &  0.00 &  IM   	\\ 
\object{J11052472-7626209}	& M2.75	&  10.5 & 1.4	&     0.25&  0.17 &  	 &  0.00 &  IM   	\\ 
\object{J11072022-7738111}	& M4.25	&  7.28 & 4.1	&  $-$0.65&  0.17 &  *	 &  0.65 &  IM   	\\ 
\object{J11101153-7733521}	& M4.5	&  10.8 & 3.4	&  $-$0.72&  0.66 &  *	 &  0.00 &  IM   	\\ 
\object{[LES2004] ChaI 423}	& M5	&  11.6 & 4.4	&  $-$0.37&  0.38 &  	 &  0.00 &  IM   	\\ 
\object{Ass Cha T2-4}		& M0.5	&  8.63 & 1.1 	&  $-$0.55&  0.25 &  *	 &  0.07 &  IM,RV	\\ 
\object{Ass Cha T2-8}		& K4	&  7.31 & 0.00	&  $-$0.53&  0.26 &  *	 &  0.05 &  IM,RV	\\ 
\object{Ass Cha T2-10}		& M3.75	&  10.9 & 1.2	&     0.05&  0.43 &  *	 &  0.00 &  IM,RV	\\ 
\object{Ass Cha T2-12}		& M4.5	&  10.5 & 0.00	&     0.03&  0.17 &  	 &  0.00 &  IM,RV	\\ 
\object{Ass Cha T2-20}		& M1.5	&  9.34 & 0.76 	&     0.26&  0.25 &  *	 &  0.00 &  IM,RV	\\ 
\object{Ass Cha T2-23}		& M4.25	&  10.0 & 0.23 	&  $-$3.19&  0.27 &  *	 &  1.0  &  IM   	\\ 
\object{Ass Cha T2-24}		& M0.5	&  9.38 & 1.8 	&     0.21&  0.28 &  	 &  0.00 &  IM,RV	\\ 
\object{Ass Cha T2-25}		& M2.5	&  9.77 & 2.3 	&     0.06&  0.17 &  	 &  0.00 &  IM,RV	\\ 
\object{Ass Cha T2-28}		& M0  	&  8.26 & 8.2  	&  $-$1.24&  0.26 &  	 &  0.87 &  IM   	\\ 
\object{Ass Cha T2-30}		& M2.5	&  9.89 & 4.5 	&  $-$1.59&  0.35 &  *	 &  0.77 &  IM   	\\ 
\object{Ass Cha T2-34}		& M3.75	&  10.0 & 2.1 	&     0.25&  0.35 &  	 &  0.00 &  IM,RV	\\ 
\object{Ass Cha T2-35}		& K8	&  9.11 & 4.0 	&  $-$1.90&  0.21 &  *	 &  1.0  &  IM,RV	\\ 
\object{Ass Cha T2-37}		& M5.25	&  11.3 & 0.67	&  $-$0.41&  0.26 &  	 &  0.00 &  IM   	\\ 
\object{Ass Cha T2-38}		& M0.5	&  9.46 & 1.3 	&  $-$1.22&  0.22 &  *	 &  0.97 &  IM,RV	\\ 
\object{Ass Cha T2-40}		& K7	&  8.24 & 2.0 	&  $-$2.40&  0.17 &  *	 &  1.0  &  IM,RV	\\ 
\object{Ass Cha T2-44}		& K5	&  6.08 & 4.7 	&  $-$5.08&  0.23 &  *	 &  1.0  &  IM,RV	\\ 
\object{Ass Cha T2-47}		& M2	&  9.18 & 3.0 	&  $-$2.93&  0.18 &  *	 &  1.0  &  IM,RV	\\ 
\object{Ass Cha T2-48}		& M3.75	&  10.0 & 0.72 	&  $-$5.73&  0.31 &  	 &  1.0  &  IM   	\\ 
\object{Ass Cha T2-49}		& M2	&  8.87 & 0.95 	&  $-$1.25&  0.18 &  	 &  1.0  &  IM,RV	\\ 
\object{Ass Cha T2-50}		& M5	&  9.84 & 0.87 	&  $-$0.13&  0.34 &  *	 &  0.00 &  IM,RV	\\ 
\object{Ass Cha T2-53}		& M1  	&  9.13 & 1.9   &  $-$0.31&  0.34 &  	 &  0.00 &  IM   	\\ 
\object{Ass Cha T2-56}		& M0.5	&  9.23 & 0.1  	&  $-$0.83&  0.20 &  *	 &  0.69 &  IM,RV	\\
\hline
\end{tabular}
\tablefoot{
\tablefoottext{a}{Identifiers recognized by the Simbad Database. Names starting with ``J'' are 2MASS identifiers \citep{cut03}.}
\tablefoottext{b}{\citet{luh04,luh07}}
\tablefoottext{c}{From the 2MASS Point Source Catalog.}
\tablefoottext{d}{Negative equivalent width values indicate lines in emission.}
\tablefoottext{e}{Equivalent widths marked with an asterisk have been corrected for absorption of a telluric standard with spectral type K7 (see Sect.~\ref{sec:datareduction}). }
\tablefoottext{f}{Probability for significant emission (see Sect.~\ref{sec:EWs})}
\tablefoottext{g}{Target was observed as part of radial velocity (RV; \citealt{ngu12}) and/or imaging (IM; \citealt{laf08}) survey and no companions were found.}
}
\end{table*}

\subsection{Observations\label{sec:obs}}
K-band spectra were acquired for each target star using the SOFI spectrograph on the ESO/NTT at La Silla Observatory. A 0.6$\times$290\arcsec\ slit was used along with a high-resolution 3rd order grism (R\,$\approx$\,2200) resulting in a full wavelength range of $\sim$1.95--2.42\,$\mu$m. To enable sky removal, both the science and calibration targets were observed in a ABBA nodding pattern with 4--8 exposures per target and exposure times of 60 to 120 seconds. The resulting S/N was typically $\gtrsim$100 per pixel at 2.2\,$\mu$m. 

Within one hour of every science observation, {a telluric standard star of late spectral type was observed (one of PMI11203-7828, K7, $T_\mathrm{eff}=4000$\,K or PMI11349-7436, M2, $T_\mathrm{eff}=3520$\,K).} Late-type stars were selected as telluric standards as they {show little or no absorption at Br$\gamma$} and will therefore facilitate removal of telluric features without affecting our equivalent width (EW) measurement. 

\subsection{Data reduction}\label{sec:datareduction}
Data reduction was done with standard procedures in IDL and IRAF and custom routines in Python. Detector crosstalk was removed using the algorithm from a routine for the SOFI spectrograph presented on the SOFI instrument webpages\footnote{\url{http://www.eso.org}}. Bad pixels and cosmic rays were removed by searching for pixels that were 4 standard deviations above or below their surroundings and replaced by the median of the nearest good pixels. The 2D spectra were then calibrated using a lamp flat and the nod pairs were subtracted from each other. {Since geometrical distortion across the field of view appears to be small -- spectral traces are aligned with the columns of the detector to within 1 pixel from end to end -- we do not apply a correction for geometrical distortion. To test whether distortion has an effect on the extracted spectra and equivalent width measurements, we compared results with and without distortion correction. No significant difference could be measured. We accordingly do not apply a distortion correction to minimize noise contributions from interpolation.} 
The spectral extraction was completed using the \textit{apextract} module in IRAF which traced and removed any remaining background. The IRAF procedure \textit{identify} was used to wavelength calibrate the raw 1D spectra to a starting wavelength of 19700\AA\ and a dispersion of 4.6\AA/pixel using 13 emission lines from a Neon-Xenon lamp. Residual bad pixels were removed from the extracted spectra. The final usable wavelength range was 2.02--2.30\,$\mu$m\ due to the detector response, high noise, and stronger telluric features in the wings of the spectrum.

The telluric star spectra were shifted by up to 2 pixels in order to align with the target spectra and divided by a blackbody of the appropriate temperature (Sect.~\ref{sec:obs}). {While M-type stars have no photospheric line features at the wavelength of Br$\gamma$, the K7 star is expected to show intrinsic absorption with an equivalent width of $\sim$0.3\AA\ \citep{dae13} which has to be removed before it can be used as a telluric standard. We isolate and remove this absorption feature by reducing a subset of our K7 standard star observations using an M-type telluric standard observed at a similar time and airmass. The thus-reduced standards were used to produce a model of the Br$\gamma$ feature which is equal to 1 at every wavelength except in the absorption feature, where it follows the average flux of the reduced standards, normalized to a continuum of 1. All observations of the K-type telluric standard were divided by this model before continuing with the reduction. 
This procedure is superior to, e.g., replacing the area around Br$\gamma$ with the continuum, since there are known telluric absorption features blending into the wings of the Br$\gamma$ feature. Stars that were reduced following this procedure are highlighted with an asterisk in Table~\ref{tab1}.
}
The individual extracted science spectra were divided by the extracted and processed telluric stars. {Then, all individual spectra of a target were averaged.}

Equivalent widths were measured {according to $EW = \int (1-F_\lambda/F_c) d\lambda$} in an 80\AA-wide region centered on Br$\gamma$.
The continuum $F_\mathrm{c}$ was measured at the position of Br$\gamma$ after removing any flux in the integration region and smoothing the whole spectrum with a 50 pixel-wide boxcar kernel. This method was verified by eye for a large number of targets and always delivered a good estimate for the continuum. The measured EW values and their 1$\sigma$ uncertainties, {inferred from the continuum noise close to the Br$\gamma$ line,} are listed in Table~\ref{tab1}. For reference, the spectra around the Br$\gamma$ line are shown in Appendix~\ref{sec:appA}.

\section{Results}\label{sec:results}
\subsection{Undetected binary companions}
Not all targeted stars have been part of a spectroscopic binary survey {(see Table~\ref{tab1})} and orbital separations between $\sim$0\farcs01 and 0\farcs1 are only poorly covered by the combination of direct imaging and spectroscopy. Therefore, a fraction of the targets might have undetected binary companions. We estimate the frequency of missed companions with separations below {$\rho_\mathrm{lim}\approx 500$\,AU}, beyond which the influence on disk evolution is {assumed to be negligible}, and with masses above the Deuterium burning limit $\sim$13\,M$_\mathrm{Jup}$. {While there have been previous observations of $\rho_\mathrm{lim}$ which resulted in smaller values \citep[e.g., $\sim$40\,AU,][]{kra12b}, we find evidence for influence of binary companions at larger separations up to 500\,AU in the present study (see Sect.~\ref{sec:discussion1}) which we use as a conservative limit.}

\begin{figure}
\centering
\includegraphics[width=\columnwidth]{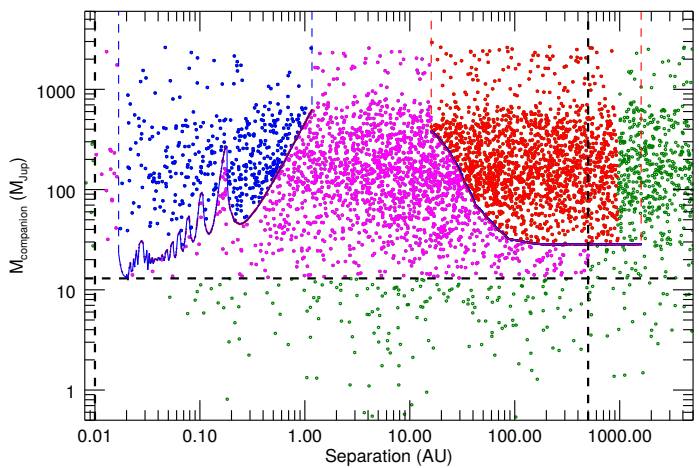}
\caption{\label{fig2}{Assessment of the missed companion fraction from a Monte Carlo simulation. Simulated companions (circles of any color) are compared with the detection limits of the RV survey \citep[left curve]{ngu12} and AO imaging survey \citep[right curve]{laf08} used for target selection. Points above these curves count as detections in our simulation (blue/left curve: RV detection, red/right curve: imaging detection). 
Companions within our survey limits (heavy dashed lines) that would remain undetected by either of these studies, i.e., in between or below their curves, are shown in magenta. Companions outside the sensitivity limits of our survey or that of \citet{ngu12} and \citet[right curve]{laf08} are shown in green.} For illustration purposes, only 10\% of all simulated points are shown.}
\end{figure}

For the {30} stars in our sample that have been part of both queried multiplicity surveys, the spectroscopic survey by \citet{ngu12} and the adaptive optics observations by \citet{laf08}, this fraction is $\sim$30\%. The rest of our sample was only part of the adaptive optics survey and expected to have undetected companions in $\sim$37\% of the cases.
This was inferred from a simulation that generates a set of 50000 companions to stars with primary masses that have been bootstrapped from our sample \citep[masses from][]{laf08}, assuming a log-normal separation distribution as observed for the field \citep{rag10} and a flat mass-ratio distribution. The latter is consistent with what has been reported for Chamaeleon\,I by \citet{laf08}. The artificial companion sample was then compared {with the sensitivity curves in \citet{laf08} and \citet{ngu12} to measure} the fraction of companions that would remain undetected by either survey (Fig.~\ref{fig2}).
The assumed method and distributions are validated by the fact that only a single overall multiplicity fraction had to be assumed to simultaneously reproduce both multiplicity fractions in the same primary mass range, i.e., 27\% detected with adaptive optics imaging and 7\% in the spectroscopic survey.

The true fraction of undetected multiples is possibly up to a few percent lower due to the existence of higher-order multiples. While the overall higher-order multiplicity frequency is on the order of 11\% over all separations and mass ratios probed by field star surveys \citep{rag10}, the frequency of wide tertiary companions to close binaries with periods $<$100\,d has been observed to be as high as 63\%$\pm$5\% \citep{tok06}. As a consequence, not all simulated companions contribute to the multiplicity frequency as they may be bound in the same systems. We will, however, keep our estimates of 30\% and 37\% for the undetected companion fraction for two reasons. Firstly, not all triple stars will have two companions within the separation and mass range of interest in the current paper, and tertiaries outside the tested region have no impact on the current statistics. The fraction of triples relevant in this context must therefore be strictly smaller than the true triple and higher-order multiple fractions. Secondly, we point out that our estimates are conservative upper limits of the multiple contamination fraction, since the existence of higher-order components effectively reduces the number of multiple stars that can be constructed with the same number of simulated companions.

We accordingly expect $\sim$19 stars in our sample to host an undetected companion within $\sim$500\,AU and above 13\,M$_\mathrm{Jup}$. This is factored into the inferred Br$\gamma$ emitter frequencies in Sect.~\ref{sec:Brgfraction}.

\subsection{Br$\gamma$ equivalent widths}\label{sec:EWs}
\begin{figure}
\centering
\includegraphics[width=\columnwidth]{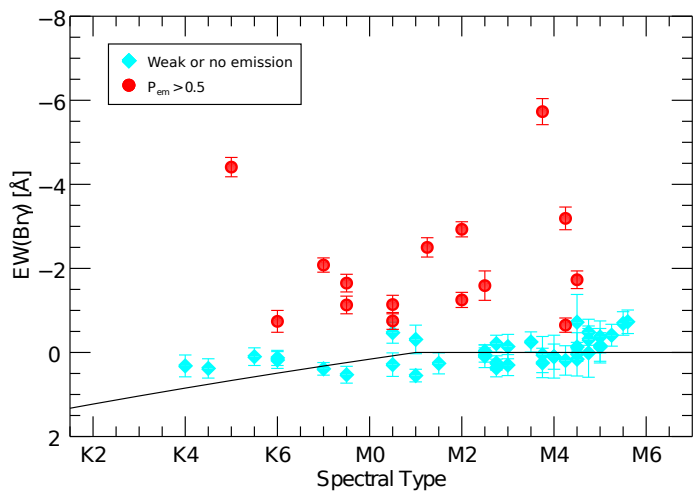}
\caption{\label{fig3}Br$\gamma$ equivalent width as a function of spectral type. The black line shows the expected photospheric value \citepalias{dae13}. Red circles are classified as significantly emitting at Br$\gamma$, blue diamonds show targets consistent with weak or no emission. }
\end{figure}

{We follow the strategy in \citetalias{dae13} to infer the frequency of Br$\gamma$ emitters among Cha\,I single stars from the number of targets with significant emission in the Br$\gamma$ line. To measure the Br$\gamma$ emitter fraction, we estimate for each target the probability $P_\mathrm{em}$ that the peak emission is significantly (3$\sigma$) above the expected photospheric value (values listed in Table~\ref{tab1}).} The latter was estimated by \citetalias{dae13} to be a function of spectral type with no detected photospheric feature from stars with spectral types equal to or later than M0 and absorption following a quadratic EW-SpT relation for earlier types. 
{Fig.~\ref{fig3} shows this relation and the resulting classification of our targets as emitters and non-emitters.}

To validate our results, we compare our Br$\gamma$ equivalent width measurements with the values derived by \citet{ant11} for a set of nine stars that our samples have in common {(CHSM\,1715, Hn\,5, ISO\,52, ISO\,237, T30, T38, T47, T49, T53)}. We find that all mutual targets that we classify as significantly emitting also have strong emission measurements by \citet{ant11}, and all but one of our non-emitters appear with weak emission in \citeauthor{ant11}. Despite this agreement it appears that their equivalent width measurements are {on average $\sim$2.5 times higher than ours for the subset of targets with significant emission.} {A similar difference has been found by \citet{bia12} when comparing their H$\alpha$ equivalent widths with those derived by \citet{ant11}. The difference might be partly due to the lower spectroscopic resolution of the \citeauthor{ant11} study. Additional effects that influence measured equivalent widths include source variability and veiling. A correction for the latter has neither been applied to our data nor to that of \citet[A. Antoniucci, \emph{priv.\ comm}]{ant11}. Veiling can accordingly not account for the difference between the two studies.}
We note that neither a systematic equivalent width offset nor veiling have an impact on our parameter of interest, i.e., the \emph{fraction} of Br$\gamma$ emitters (see Sect.~\ref{sec:Brgfraction}), as long as features are not reduced below the sensitivity of the survey. This is because of the additive nature of the veiling correction which scales the line flux in the same way as the continuum (as long as $f_\lambda\approx f_\mathrm{cont,\lambda}$) with no immediate impact on the significance of a detection. 

\subsection{Br$\gamma$ emitter fraction of singles in Cha\,I}\label{sec:Brgfraction}
{We find 15 stars with significant emission, i.e., an emitter probability $P_\mathrm{em}>0.5$ and 39 with weak or no detectable emission. This is equivalent to a raw Br$\gamma$ emitter fraction of $F_\mathrm{raw}=27^{+7}_{-6}$\%, using Bayesian statistics of the emitter probabilities in Table~\ref{tab1} following the method described in \citetalias{dae13}.} The true value is likely higher because of the low emitter fraction of stars with unseen companions and veiling, which can reduce the detected equivalent width to fall below detectability. We discuss these and other effects in the following.

One bias results from variable uncertainties of the Br$\gamma$ EW measurements. Since large uncertainties will result in a preference for a measurement to be consistent with zero emission, we exclude all stars with {a measurement uncertainty of} $\Delta$EW$>$0.5\AA\ from the statistics. This value was chosen to match the distribution cut-off for EW measurements in \citetalias{dae13} -- only two of their measurements were above this value -- and will ensure that both samples are comparable. The sample size after this cut is 51 stars, the corrected emitter fraction is $F=28.5^{+7.5}_{-6.5}$\%.

As derived in Sect.~\ref{sec:biases}, it is possible that a significant fraction of all stars have previously undetected stellar or substellar companions. Multiplying the number of stars after the $\Delta$EW cutoff (28 with AO and RV measurements and 23 with only AO) by the expected companion fractions, we expect $\sim$17 stars to have unseen companions. Since we are interested in finding the true disk frequency of \emph{single} stars, we use this number to estimate the limits of the inferred disk frequency in two ways. The first, conservative approach assumes that either all 17 potential binaries are part of the Br$\gamma$-emitting subsample, or that none of them shows Br$\gamma$ in emission. This results in lower and upper limits for the single star Br$\gamma$ emitter fraction of $F\le6.5$\% (68\% confidence) and $F=41.5^{+9.5}_{-8.5}$\%, respectively. 
The second approach exploits the fact that an estimate of the disk frequency in binary stars can be made from existing observations by \citetalias{dae13}. Reanalyzing the \citeauthor{dae13} data using the same methodology and criteria as in the current paper (spectral type range, sensitivity cutoff), and focusing only on close binaries with separations below 100\,AU which represent by far the largest population of possibly contaminating binaries in the current sample (see Fig.~\ref{fig2}), we derive a Br$\gamma$ emitter frequency of {$6.5^{+16.5}_{-3.0}$\%} for binary stars. Correcting the above-derived value of $F=28.5^{+7.5}_{-6.0}$\% for the frequency of undetected binary companions, this leads to a single star emitter fraction of $F=39.5^{+14.0}_{-9.9}$\%. This is our best estimate of the single star Br$\gamma$ emitter fraction in Chamaeleon\,I. 

For completeness, we convert the derived Br$\gamma$ emitter fractions to accretion fractions that can be compared with other surveys, e.g., those using H$\alpha$ as an accretion indicator. These are known to show a larger accretion fraction than Br$\gamma$ surveys. Applying the correction {of +15\%} derived from the fraction of classical T Tauri Stars without detected Br$\gamma$ emission \citep{fol01,dae12a} we expect another {three} targets to be classified as accreting which do not show Br$\gamma$ in emission. Including these in our calculations, we find {an accretor fraction of Cha\,I single stars of $F_\mathrm{acc}=47.8^{+14.0}_{-9.9}$\%. This is slightly higher than -- but consistent with -- previous estimates that do not explicitely exclude binaries from their samples. For example, we infer an accretor fraction in Cha\,I of 41\% from data provided by the \emph{Spitzer} and H$\alpha$ survey by \citet{dam07}. For a sample of very low-mass stars (spectral types between M5 and M9.5), \citet{moh05} find an accretor fraction of 44$\pm$8\%.}

\subsection{Br$\gamma$ emitter fraction of binaries in Cha\,I}\label{sec:binaryBrgfraction}
The largest sample of resolved binary star observations in Chaemaeleon\,I is provided by \citetalias{dae13}. We reanalyze their data with the same method and selection as presented in Sect.~\ref{sec:Brgfraction}. The resulting sample of {36 components} of binary stars with separations ranging from $\sim$20\,AU to 800\,AU reveals a Br$\gamma$ emitter frequency {of 20$^{+9}_{-6}$\%}. The fact that these numbers are different from the values quoted in the original paper is due to the different target selection to ensure comparability with the current study. As expected, we recover a strong dependence on binary separation. The Br$\gamma$ emitter fraction of close-separation binaries $<$100\,AU is {$6.5^{+16.5}_{-3.0}$\%}, while that of wider systems $100<\rho\le500$\,AU is {$38^{+14}_{-12}$\%}.

\section{Discussion}\label{sec:discussion}
\subsection{Binary stars show accretion significantly less often than singles}\label{sec:discussion1}
Our study demonstrates that the frequency of Br$\gamma$ emitters among stars in binary systems is significantly reduced compared to single stars in the same environment. As derived in Sect.~\ref{sec:Brgfraction}, our best estimate for the Br$\gamma$ emission frequency of single stars in Chamaeleon\,I is $F=39.5^{+14.0}_{-9.9}$\%. The emitter frequency of stars bound in binary systems appears to be reduced by a factor of 2 compared to this value. The effect is strongest for close binaries which appear to have a $\sim$2.4--19 times lower emitter fraction (68\% confidence interval) {while wide binaries show no significant decrease.}

As shown by \citet{ant11}, Br$\gamma$ emission correlates well with the strength of accretion and mass accretion rates in T Tauri stars. Accordingly, we can interpret the significant difference between single and binary star Br$\gamma$ emission as confirmation for significantly reduced occurence of accretion activity in binaries.
Assuming that most or all low-mass stars, irrespective of whether they are bound in a binary system, transition through the T\,Tauri phase (equivalent to a $\sim$100\% disk fraction at birth), the reduced accretor fraction can be explained through either a shorter epoch of accretion or lower mass accretion rates which fall more easily below the detection threshold. Lower accretion rates could not be observed in a comparison of singles and multiples in several star-forming regions \citep[]{whi01,dae12a}. Accordingly, a shorter accretion duration may be the cause for a lower accretion fraction, which is expected when assuming constant mass accretion rates and considering that disks in binaries are truncated to radii of 1/2 to 1/3 of the orbital semi-major axis along with a reduction in disk mass \citep{art94,har12}. In the T\,Tauri phase no significant accretion from an envelope replenishes the disk, the disk material will accordingly be accreted or dispersed earlier when disk sizes and masses are reduced.

A reduced disk lifetime has implications for giant planet formation. 
To date we know of more than {60} planets that are in orbit around components of binary stars \citep{roe12}. 
Some of these binaries have separations $<$100\,AU. This is consistent with two formation channels for these systems. 
a) The observed short accretion timescales may indicate that a part of the planet population forms on very short timescales $<$2\,Myr. This would hint at a rapid formation process, such as disk fragmentation. Accordingly, planets formed through a slower process should be preferably found in systems with longer disk evolution timescales, i.e., wider binaries or single stars. Evidence for this dichotomy has been presented by \citet{duc10}, who investigated the occurence of planets in binary stars of various separations. They find that systems closer than $\lesssim$100\,AU only harbor massive planets $>1\,M_\mathrm{Jup}$, while wider systems can have any mass above or below this threshold. 
b) Alternatively, dynamical evolution may have altered the orbit of the stellar system only after the planet has formed and migrated to its final position. Such dynamical orbit evolution at later stages of star formation up to $\sim$100\,Myr are in agreement with dynamical studies and the fact that the peak of the binary star separation distribution may change as a function of age \citep{rei12,lei93}. This formation branch of close binaries with planets is counter-acted by the fact that dynamical evolution preferably removes light companions, i.e., planets, from multiple systems due to their low binding energy. This would explain the low frequency of planets in close visual binary systems, in agreement with the reduced multiplicity fraction of close binaries \citep{egg10}. It remains to be investigated whether the observed frequency of companions in close binary systems agrees with a purely dynamical evolution scenario. 

It is possible that both channels contribute to the formation of the observed binary systems with and without planets but observational biases and the small number of known systems currently inhibit strong conclusions about which might be the dominant mechanism. The significantly reduced accretion frequency in binaries found in the present study, however, suggests that the gas giant planet formation timescales may be well-matched to the disk evolution timescale {which would conveniently explain the observed reduced planet frequency in close binaries.}

\section{Conclusions}\label{sec:conclusion}
The current paper presents a measurement of the protoplanetary disk frequency around single stars in the 2--3\,Myr-old Chamaeleon\,I star-forming region, for the first time taking into account the significant fraction of undetected binary stars in a typical target star sample. We use spectroscopic observations at 2.2\,$\mu$m\ of 54 stars to investigate the frequency of stars with emission in the Brackett-$\gamma$ line, a feature sensitive to ongoing accretion from a protoplanetary disk. To minimize the possibility of undetected binarity within $\sim$100\,AU, targets were selected from a high-angular resolution survey and matched with radial velocity surveys. We further use simulations to estimate the chance of undetected multiplicity to be 30--37\% which is used to infer the true fraction of emitters among single stars in Cha\,I. We compare the results with binary stars in the same region which were previously investigated using the same accretion indicator and sample structure \citep{dae13}.

We conclude the following:
\begin{enumerate}
  \item The best estimate of the Br$\gamma$ emitter frequency of single stars in Chamaeleon\,I is $39.5^{+14.0}_{-9.9}$\%. 
  \item {Using the same methods to reanalyze literature data of binary star components in Chamaeleon\,I, we find that the frequency of accretors bound in close binaries with separations $<$100\,AU is lower than that of single stars by a factor of 2.4--19 (68\% confidence). The frequency of Br$\gamma$ emitters in wider binaries appears to be consistent with that of single stars in the same region.} Owing to the coherence of the analysis, this result represents one of the most robust confirmations that the accretion disk frequency of binary stars is reduced compared to single stars in the same star-forming region. 
  \item Based on a sensitivity correction using the frequency of Br$\gamma$ emitters among classical T Tauri stars, we convert our Br$\gamma$ emitter frequency to an accretor fraction of Chamaeleon\,I singles of $47.8^{+14.0}_{-9.9}$\%. This is slightly higher than, but consistent with, previous accretion frequency measurements in the region.
\end{enumerate}

We conclude that the influence of multiplicity on protoplanetary disk evolution is significant. Accordingly, a clear separation of single and binary stars must be applied to any observational study of protoplanetary disks in order to serve as input for theoretical studies. More surveys like the present are needed to homogeneously analyze the evolution of disks, both around single and binary stars, as a function of age.

\begin{acknowledgements}
We thank the unknown referee for valuable comments and suggestions. 
We thank Veselin Kostov for feedback on the manuscript and Simone Antoniucci for detailed information about their previous measurements.
This work was supported by NSERC grants to RJ. SD is supported by a McLean postdoctoral fellowship. 
This publication makes use of data products from the Two Micron All Sky Survey, which is a joint project of the University of Massachusetts and the Infrared Processing and Analysis Center/California Institute of Technology, funded by the National Aeronautics and Space Administration and the National Science Foundation.
This research has made use of the SIMBAD database, operated at CDS, Strasbourg, France 
\end{acknowledgements}

\appendix
\section{Spectra}\label{sec:appA}
\begin{figure*}
\centering
\includegraphics[width=\textwidth]{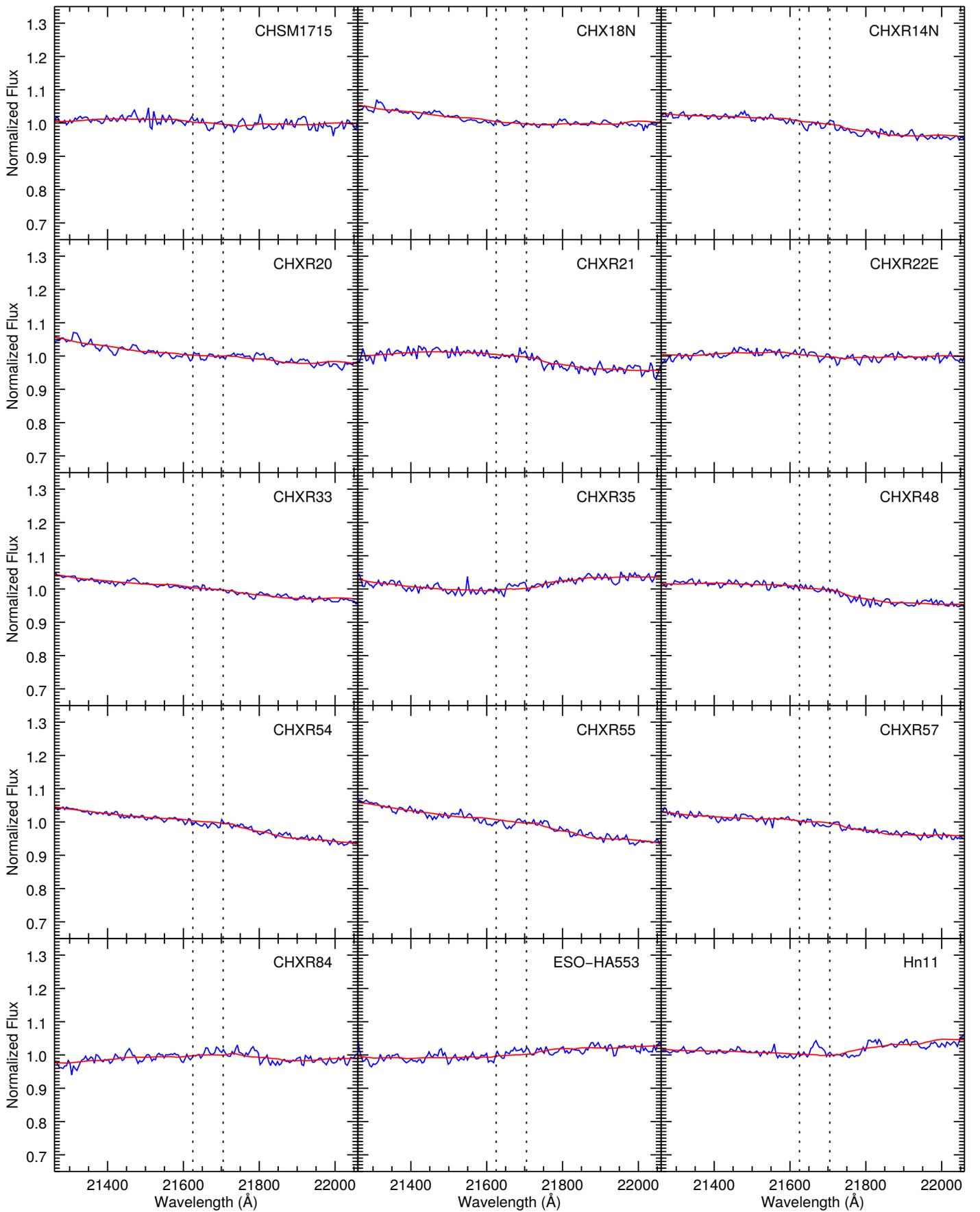}
\caption{\label{fig:A1}Spectra reduced with an M-type telluric standard are shown in blue. Our continuum estimate is overplot in red, the integration region to calculate $W_{\mathrm{Br}\gamma}$ is shown with dotted lines.}
\end{figure*}

Fig.~\ref{fig:A1} shows spectra of all 54 targets in the wavelength region around Br$\gamma$.

\addtocounter{figure}{-1}
\begin{figure*}
\centering
\includegraphics[width=\textwidth]{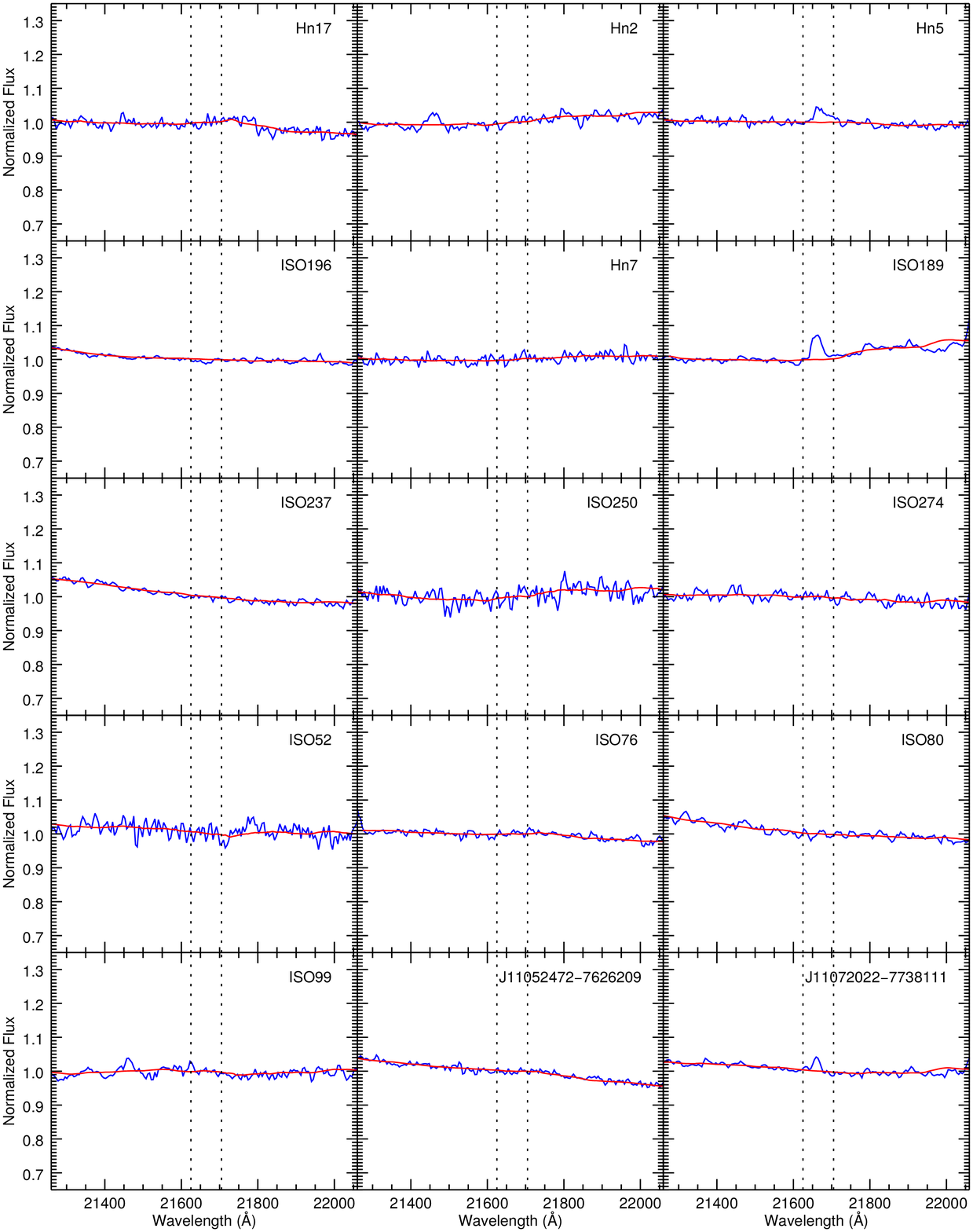}
\caption{\emph{ctd...}}
\end{figure*}
\addtocounter{figure}{-1}
\begin{figure*}
\centering
\includegraphics[width=\textwidth]{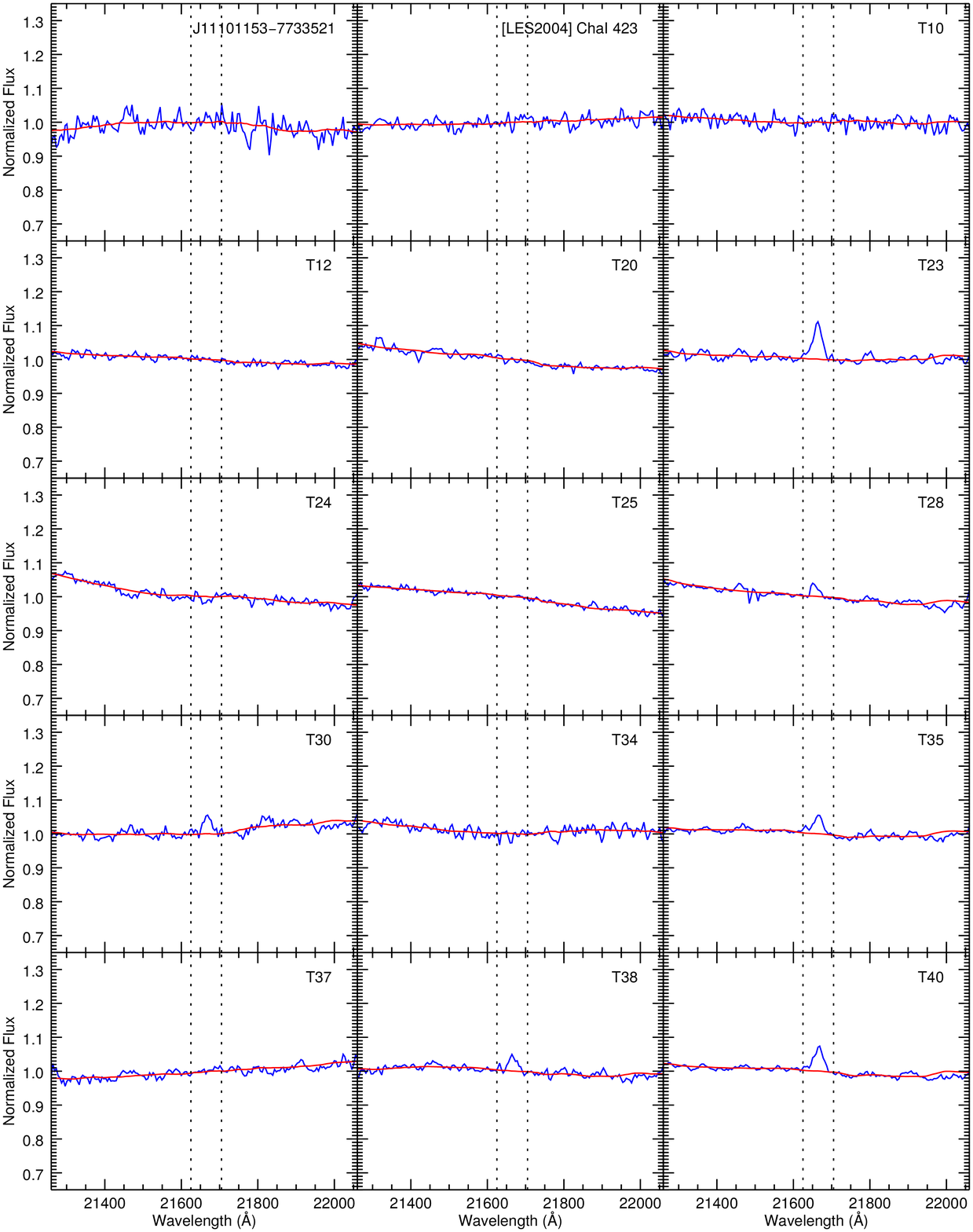}
\caption{\emph{ctd...}}
\end{figure*}
\addtocounter{figure}{-1}
\begin{figure*}
\centering
\includegraphics[width=\textwidth]{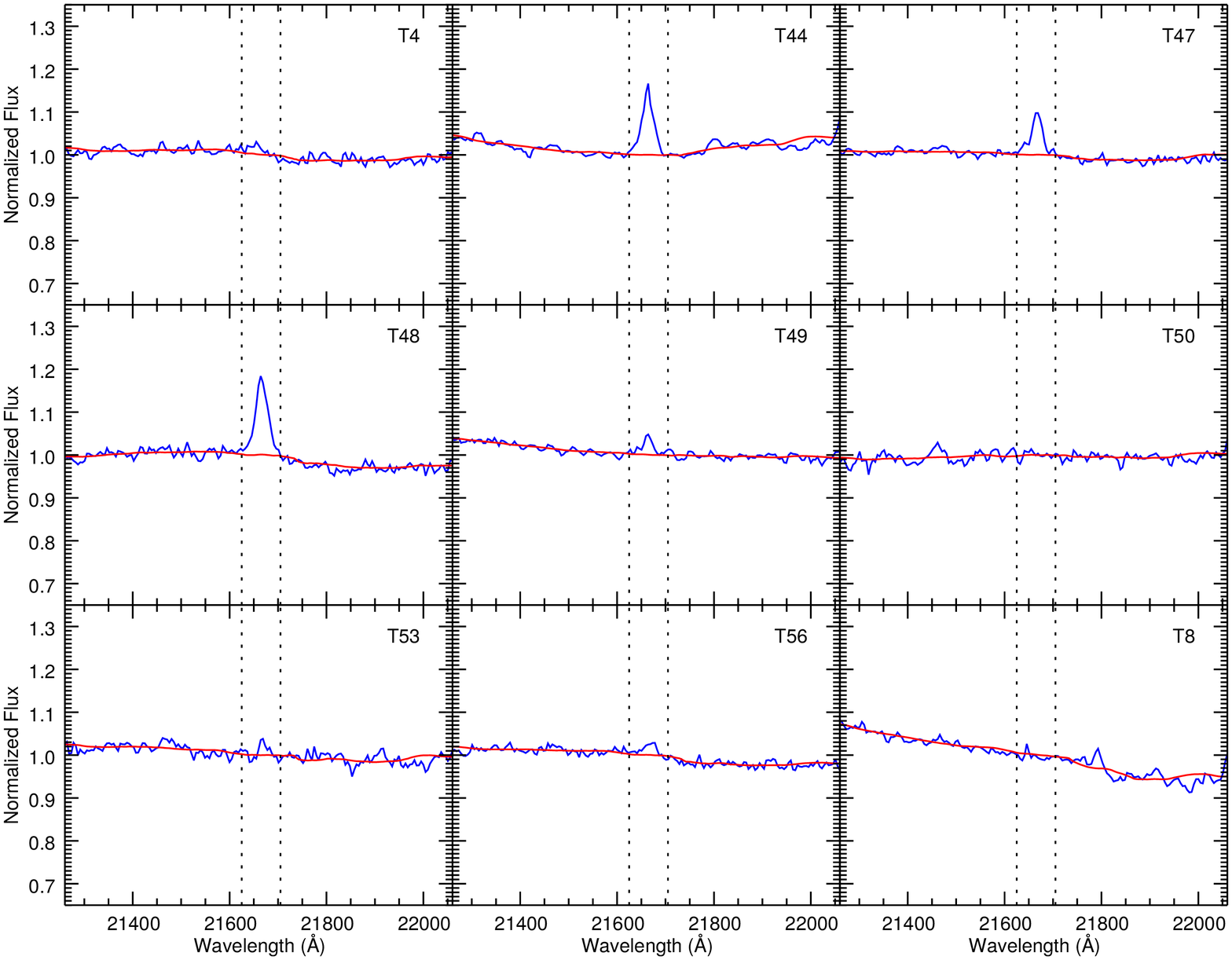}
\caption{\emph{ctd...}}
\end{figure*}

\clearpage


\begin{thebibliography}{30}
\expandafter\ifx\csname natexlab\endcsname\relax\def\natexlab#1{#1}\fi

\bibitem[{Alexander \& Armitage(2009)}]{ale09}
Alexander, R.~D. \& Armitage, P.~J. 2009, ApJ, 704, 989

\bibitem[{Antoniucci {et~al.}(2011)Antoniucci, Garc\'ia~L\'opez, Nisini,
  Giannini, Lorenzetti, Eisl\"offel, Bacciotti, Cabrit, Caratti~o Garatti,
  Dougados, \& Ray}]{ant11}
Antoniucci, S., Garc\'ia~L\'opez, R., Nisini, B., {et~al.} 2011, A\&A, 534, 32

\bibitem[{Artymowicz \& Lubow(1994)}]{art94}
Artymowicz, P. \& Lubow, S.~H. 1994, ApJ, 421, 651

\bibitem[Biazzo et al.(2012)]{bia12} Biazzo, K., Alcal{\'a}, J.~M., Covino, E., et al.\ 2012, \aap, 547, A104

\bibitem[{Boss(1997)}]{bos97}
Boss, A.~P. 1997, Science, 276, 1836

\bibitem[{Cutri {et~al.}(2003)Cutri, Skrutskie, van Dyk, Beichman, Carpenter,
  Chester, Cambresy, Evans, Fowler, Gizis, Howard, Huchra, Jarrett, Kopan,
  Kirkpatrick, Light, Marsh, {McCallon}, Schneider, Stiening, Sykes, Weinberg,
  Wheaton, Wheelock, \& Zacarias}]{cut03}
Cutri, R.~M., Skrutskie, M.~F., van Dyk, S., {et~al.} 2003, {VizieR} Online
  Data Catalog, 2246

\bibitem[{Daemgen {et~al.}(2012)Daemgen, Correia, \& Petr-Gotzens}]{dae12a}
Daemgen, S., Correia, S., \& Petr-Gotzens, M.~G. 2012, A\&A, 540, 46

\bibitem[{Daemgen {et~al.}(2013)Daemgen, Petr-Gotzens, Correia, Teixeira,
  Brandner, Kley, \& Zinnecker}]{dae13}
Daemgen, S., Petr-Gotzens, M.~G., Correia, S., {et~al.} 2013, A\&A, 554, 43

\bibitem[Damjanov et al.(2007)]{dam07} Damjanov, I., 
Jayawardhana, R., Scholz, A., et al.\ 2007, \apj, 670, 1337

\bibitem[{Duch\^ene(2010)}]{duc10}
Duch\^ene, G. 2010, ApJ, 709, L114

\bibitem[{Eggenberger \& Udry(2010)}]{egg10}
Eggenberger, A. \& Udry, S. 2010, in Planets in Binary Star Systems, ed: N.
  Haghighipour, 1st edn., Vol. 366 (Springer, Berlin), 19

\bibitem[{Fedele {et~al.}(2010)Fedele, van~den Ancker, Henning, Jayawardhana,
  \& Oliveira}]{fed10}
Fedele, D., van~den Ancker, M.~E., Henning, T., Jayawardhana, R., \& Oliveira,
  J.~M. 2010, A\&A, 510, 72

\bibitem[Folha \& Emerson(2001)]{fol01}
Folha, D.~F.~M., \& Emerson, J.~P.\ 2001, \aap, 365, 90 

\bibitem[{Harris {et~al.}(2012)Harris, Andrews, Wilner, \& Kraus}]{har12}
Harris, R.~J., Andrews, S.~M., Wilner, D.~J., \& Kraus, A.~L. 2012, ApJ, 751,
  115

\bibitem[{Jayawardhana {et~al.}(2006)Jayawardhana, Coffey, Scholz, Brandeker,
  \& van Kerkwijk}]{jay06}
Jayawardhana, R., Coffey, J., Scholz, A., Brandeker, A., \& van Kerkwijk, M.~H.
  2006, ApJ, 648, 1206

\bibitem[Joergens(2008)]{joe08} Joergens, V.\ 2008, \aap, 492, 545 

\bibitem[{Kraus {et~al.}(2012)Kraus, Calvet, Hartmann, Hofmann, Kreplin,
  Monnier, \& Weigelt}]{kra12b}
Kraus, S., Calvet, N., Hartmann, L., {et~al.} 2012

\bibitem[{Lafreni{\`e}re {et~al.}(2008)Lafreni{\`e}re, Jayawardhana, \& van
  Kerkwijk}]{laf08}
Lafreni{\`e}re, D., Jayawardhana, R., \& van Kerkwijk, M.~H. 2008, ApJ, 689,
  L153

\bibitem[{Leinert {et~al.}(1993)Leinert, Zinnecker, Weitzel, Christou, Ridgway,
  Jameson, Haas, \& Lenzen}]{lei93}
Leinert, C., Zinnecker, H., Weitzel, N., {et~al.} 1993, A\&A, 278, 129

\bibitem[{Luhman(2004)}]{luh04}
Luhman, K.~L. 2004, ApJ, 602, 816

\bibitem[{Luhman(2007)}]{luh07}
Luhman, K.~L. 2007, ApJSS, 173, 104

\bibitem[Luhman et al.(2008)]{luh08b} Luhman, K.~L., Allen, 
L.~E., Allen, P.~R., et al.\ 2008, \apj, 675, 1375 

\bibitem[{Mohanty {et~al.}(2005)Mohanty, Jayawardhana, \& Basri}]{moh05}
Mohanty, S., Jayawardhana, R., \& Basri, G. 2005, ApJ, 626, 498

\bibitem[{Monin {et~al.}(2007)Monin, Clarke, Prato, \& {McCabe}}]{mon07}
Monin, J.-L., Clarke, C.~J., Prato, L., \& {McCabe}, C. 2007, Protostars and
  Planets V, 395

\bibitem[{Nguyen {et~al.}(2012)Nguyen, Brandeker, van Kerkwijk, \&
  Jayawardhana}]{ngu12}
Nguyen, D.~C., Brandeker, A., van Kerkwijk, M.~H., \& Jayawardhana, R. 2012,
  ApJ, 745, 119

\bibitem[{Prato {et~al.}(2003)Prato, Greene, \& Simon}]{pra03}
Prato, L., Greene, T.~P., \& Simon, M. 2003, ApJ, 584, 853

\bibitem[{Raghavan {et~al.}(2010)Raghavan, {McAlister}, Henry, Latham, Marcy,
  Mason, Gies, White, \& ten Brummelaar}]{rag10}
Raghavan, D., {McAlister}, H.~A., Henry, T.~J., {et~al.} 2010, ApJSS, 190, 1

\bibitem[{Reipurth \& Mikkola(2012)}]{rei12}
Reipurth, B. \& Mikkola, S. 2012, Nature, 492, 221

\bibitem[{Roell {et~al.}(2012)Roell, Neuh\"auser, Seifahrt, \&
  Mugrauer}]{roe12}
Roell, T., Neuh\"auser, R., Seifahrt, A., \& Mugrauer, M. 2012, A\&A, 542, 92

\bibitem[Schmidt et~al.(2013)]{sch13} Schmidt, T.~O.~B., Vogt, N., Neuh{\"a}user, R., Bedalov, A., \& Roell, T.\ 2013, \aap, 557, A80 

\bibitem[{Tokovinin {et~al.}(2006)Tokovinin, Thomas, Sterzik, \& Udry}]{tok06}
Tokovinin, A., Thomas, S., Sterzik, M., \& Udry, S. 2006, A\&A, 450, 681

\bibitem[{White \& Ghez(2001)}]{whi01}
White, R.~J. \& Ghez, A.~M. 2001, ApJ, 556, 265

\bibitem[{Whittet {et~al.}(1997)Whittet, Prusti, Franco, Gerakines, Kilkenny,
  Larson, \& Wesselius}]{whi97}
Whittet, D. C.~B., Prusti, T., Franco, G. A.~P., {et~al.} 1997, A\&A, 327, 1194

\end{thebibliography}
\end{document}